\documentclass[11pt,a4paper]{article}
\usepackage{amsmath,amsfonts}
\makeatletter \@addtoreset{equation}{section}

\pdfoutput=1
\usepackage[pdftex]{graphics}
\usepackage{cite}
\usepackage{bm}
\usepackage{dcolumn}
\usepackage{graphicx}
\usepackage{subfig}
\usepackage{float}
\makeatother
\def\one{{\hbox{ 1\kern-.8mm l}}}
\newcommand{\Dslash}{\not{\hbox{\kern-4pt $D$}}}
\newcommand{\pdslash}{\not{\hbox{\kern-2pt $\partial$}}}
\newcommand{\be}{\begin{equation}}
\newcommand{\bea}{\begin{eqnarray}}
\newcommand{\eea}{\end{eqnarray}}
\newcommand{\ba}{\begin{array}}
\newcommand{\ea}{\end{array}}
\newcommand{\ee}{\end{equation}}

\newcommand{\bss}{\begin{split}}
\newcommand{\ess}{\end{split}}

\begin{document}

\begin{titlepage}
\vspace*{1mm}%
\hfill%
\vbox{
    \halign{#\hfil        \cr
           IPM/P-2011/017 \cr
                     } 
      }  
\vspace*{15mm}%
\begin{center}

{{\Large {\bf On Holography of Julia-Zee Dyon }}}

\vspace*{15mm} \vspace*{1mm}{Davood Allahbakhshi}

 \vspace*{1cm}

{\it Department of Physics, Sharif University of Technology \\
P.O. Box 11365-9161, Tehran, Iran}

\vspace*{.4cm}

{\it School of physics, Institute for Research in Fundamental Sciences (IPM)\\
P.O. Box 19395-5531, Tehran, Iran \\}

\vspace*{.4cm}

\it{allahbakhshi@ipm.ir}

\vspace*{2cm}
\end{center}

\begin{abstract}
The holographic dual of self gravitating Julia-Zee dyon is
discussed. It is shown that the dual field theory is generally a
field theory with a vortex condensate. The vacuum expectation
values of the dual operators, as functions of the sources in the
field theory, are studied in a class of bulk solutions. In these
solutions the sign of the vacuum expectation values of the dual
operators change, by changing the sources in the model.
\end{abstract}

\end{titlepage}

\section{Introduction}

It is well known that the AdS/CFT correspondence
\cite{Maldacena:1997re,Witten:1998qj,Gubser:1998bc}, is a useful
tool in understanding at least the qualitative features of
strongly coupled field theories. During past years many
holographic models are studied. In some cases the bulk theories
are constructed so as to simulate certain features of a known
quantum field theory, e.g., AdS/QCD and AdS/CMT models; and in
some cases a field theory is proposed as the dual theory of a
given bulk theory. In both approaches, by the well known recipe of
AdS/CFT correspondence, the key considerations are symmetries and
the field contents of the theories. In some cases the comparable
quantities and behaviors are illuminating for finding the dual
theories.

In this work we study one of the gravitational systems recently
studied, the \emph{self gravitating Julia-Zee dyon}
\cite{Lugo:2010qq,Lugo:2010ch,Allahbakhshi:2010ii}. A.R.Lugo,
E.F.Moreno and F.A.Schaponsik, have found a numerical solutions
for the Julia-Zee black dyon, in the BPS limit,
with\cite{Lugo:2010ch} and without\cite{Lugo:2010qq} back reaction
on the metric, and they found that the vacuum expectation values
vanish by raising the temperature of the bulk blackhole. It can be
considered as a phase transition from a dyonic blackhole. In
\cite{Allahbakhshi:2010ii}, it is shown that a simple, vanishing
scalar solution for the dyon, becomes dynamically unstable by
lowering the temperature of the bulk blackhole to a scalar hairy
blackhole, which means that the dual field theory changes its
vacuum, so by lowering the temperature of the dual boundary
system, a phase transition occurs to a scalar condensed phase.
Multimonopole solutions with AdS background are studied with axial
\cite{vanderBij:2002sq,Radu:2004ys} and crystalline
\cite{Bolognesi:2010nb,Sutcliffe:2011sr} symmetry.
In\cite{Bolognesi:2010nb}, D.Tong and S.Bolognesi have studied the
multi monopole solution at the BPS limit, and found that at low
temperatures, the multi monopole solutions are favored in
comparison to blackhole single monopole solutions. As the multi
monopole solutions in AdS background can have crystalline
symmetry, they have proposed that these configurations could
correspond to strange metals at low temperatures. In this article
we study the holographic vacuum expectation values for the dual
operators, by the well known recipe of the Gauge/Gravity duality;
and also calculate a numerical solution to the equations of
motion, to determine the behavior of the v.e.vs of the dual
operators, as functions of the sources.

The paper is organized as follows. In the next section we
introduce the bulk action, equations of motion and the form of the
Julia-Zee dyon. In section (3) we calculate the vacuum expectation
values of the dual operators, generally from
Einstein-Yang-Mills-Higgs action and as a special case, for
Julia-Zee dyon. In section (4) we find a set of numerical
solutions for the dyon, and generically study their thermodynamics
and v.e.v of the dual operators. In the last section we end with
conclusion.

\section{The model}
The action we consider in this article is
\cite{Allahbakhshi:2010ii}: \be S \ = \int \sqrt{-g}\; d^4x\
\big[\frac{1}{16 \pi G}(R+\frac{6}{L^2})-\frac{1}{4}F^{a}_{\mu
\nu}F^{a\mu\nu}-\frac{1}{2}D_\mu\phi ^a D^\mu \phi ^a
-\frac{\lambda}{4}(\phi^a\phi^a)^2+
 \frac{1}{L^2}(\phi^a\phi^a) \big], \ee
The covariant derivative and the field strength are:

\be D_\mu\phi^a=\partial_\mu \phi^a + e \epsilon^{abc}A^b_\mu
\phi^c\;;\;F^a_{\mu\nu} = \partial_\mu A^a_\nu - \partial_\nu
A^a_\mu+ e \epsilon^{abc}A^b_\mu A^c_\nu. \ee
As is clear, the mass term for the scalar field is chosen to be $m^2=-2/L^2$. Since the su(2) gauge symmetry of the bulk action corresponds to a global su(2) symmetry in dual field theory, so the scalar field can be considered as the dual of a \emph{two fermion flavored composite operator} (similar to pions or a flavored Cooper pair). In this sense, since the mass dimension of a fermion in 2+1 dimensions is \emph{one}, the operator dual to the $\phi$, in the field theory, will have mass dimension \emph{two}. On the other hand the source that couples to such an operator has mass dimension \emph{one}\footnote{It is similar to a Dirac or Majorana mass.}. In order to have these special mass dimensions, the asymptotic form of the bulk scalar field should be:
\be
\phi=\frac{\tilde{\phi}_0}{r}+\frac{\tilde{\phi}_1}{r^2}+...\;.
\ee
So $m^2$, has to be $-2/L^2$.

The equations of motion are: \be \begin{split}
&R_{\mu\nu}-\frac{1}{2}(R+\frac{6}{L^2})\;g_{\mu\nu}=8\pi G
\;T_{\mu\nu}\cr &T_{\mu\nu}=\big[-\frac{1}{4}F_{\rho\lambda}^a
F^{a\rho\lambda}-\frac{1}{2}D_\lambda\phi^a D^\lambda\phi^a-
V(\phi^2) \big]
g_{\mu\nu}+F^a_{\mu\lambda}F^{a\lambda}_\nu+D_\mu\phi^a
D_\nu\phi^a\cr\cr &D_\mu\big( \sqrt{-g}D^\mu\phi
\big)^a-\sqrt{-g}\;\frac{\delta V}{\delta \phi^a}=0 \cr\cr
&D_\mu\big( \sqrt{-g}F^{\mu\nu}
\big)^a-e\sqrt{-g}\;\epsilon^{abc}\;\phi^b\;(D^\nu\phi)^c=0
\end{split} \ee
The Julia-Zee ansatz has the form:
\bea\begin{split}\phi=\vec{\sigma}.\vec{\phi}&=\frac{H(r)}{er}\;
\sigma^r \cr A=\vec{\sigma}.\vec{A} &=\frac{J(r)}{er}\;\sigma^r
\;dt + \frac{1-K(r)}{e}\;\big(-\sigma^\varphi \; d\theta +
sin(\theta)\;\sigma^\theta \; d\varphi \big),
\end{split}\eea
where $\sigma^r,\sigma^\theta,\sigma^\varphi$ are defined as:\be
\sigma^r=\hat{r}.\vec{\sigma}\; ,
\;\sigma^\theta=\hat{\theta}.\vec{\sigma}\; , \;
\sigma^\varphi=\hat{\varphi}.\vec{\sigma} \ee and $\hat{r}\; ,
\;\hat{\theta}\; , \;\hat{\varphi}$ are the usual unit vectors of
spherical coordinates \footnote{we name this frame as spherical
frame of su(2).}. We consider the metric in the most general form
of a spherical symmetric metric in global coordinates:\be ds^2=\;
-e^{X(r)}\;dt^2\;+\;e^{Y(r)}\;dr^2\;+\;r^2\;d\Omega_2^2 \ee

There are two simple solutions for the full equations of motion
\cite{Kasuya:1981tq,Allahbakhshi:2010ii}: \bea \begin{split} H&=C
r\;,\; K=0\;,\;J=\mu r -
\rho\;,\;e^X=e^{-Y}=1-\frac{2M}{r}+\frac{q^2}{r^2}+\frac{r^2}{\tilde{L}^2}\cr
H&=0\;,\; K=0\;,\;J=\mu r -
\rho\;,\;e^X=e^{-Y}=1-\frac{2M}{r}+\frac{q^2}{r^2}+\frac{r^2}{L^2}
\end{split}\eea where $C=\sqrt{2/\lambda L^2}$ is the minimum of
the potential, and q is related to $\rho$ through:\be
q^2=\frac{4\pi G_N(1+\rho^2)}{e^2}, \ee and $\rho$ is related to
the chemical potential by \be \rho=\mu r_H. \ee $\tilde{L}$
satisfies:\be \frac{1}{\tilde{L}^2}=\frac{1}{L^2}+\frac{8\pi
G_N}{3\lambda L^4}.\ee It is very easy to see that for the above
solution there is a relation between the temperature of the
blackhole and chemical potential as:\be T+\alpha\; \mu^2=\beta \ee
where $\alpha$ and $\beta$ are two constants related to the
parameters of the solution ($\lambda , r_H , G_N , e , L$). This
relation is in fact the first law of thermodynamics: \be
\frac{\epsilon+P}{s}=T+\mu \; \frac{\rho}{s}\;;\;\rho\propto\mu\ee

\section{Vacuum of dual theory}
To understand the vacuum of the dual field theory, we should
calculate the vacuum expectation values of the operators in the
field theory. In the theory under consideration we have a scalar
field, $\phi^a$, which is dual to a scalar operator, $O^a$, and a
gauge field $A^a_\mu$, which corresponds to a vector operator,
$V^a_\mu$, in the dual field theory.

Before we proceed to calculate the one point functions of the
operators, we recall the linear method for calculating them in
AdS/CFT.

Suppose that we have a bulk theory which is defined by an action
S($\phi$). In order to calculate the 1-point functions we need to
vary the on-shell action with respect to the fields. Expanding to the second order we
have:\bea\begin{split}
S_{on-shell}(\phi_B+\delta\phi)&=S^{(0)}(\phi_B)+S^{(1)}(\phi_B,\delta\phi)+S^{(2)}(\phi_B,\delta\phi^2)+...;\end{split}\eea
the first term is the free energy of the dual theory. The second
and third terms can be written in the
forms:\bea\begin{split}S^{(1)}&=\int{\delta\phi.(\text{b.e.m})}+\;S_N^{(1)}(\phi_B,\delta\phi)\cr
S^{(2)}&=\int{\delta\phi.(\text{f.e.m})}+\;S_N^{(2)}(\phi_B,\delta\phi^2)\end{split}\eea
the terms b.e.m and f.e.m refer to the expressions for the
background equation of motion and that of the linearized equation
of motion for the fluctuation, respectively; which are zero. Now
suppose that the asymptotic expansion of the fields are:\bea
\begin{split}\phi_B&=\tilde{\phi}_s
r^{\Delta_+}+...+\tilde{\phi}_c
r^{\Delta_-}+...,\cr\delta\phi&=\delta\tilde{\phi}_p
r^{\Delta_+}+...+\delta\tilde{\phi}_r
r^{\Delta_-}+...,\end{split}\eea where the subscripts
\emph{s,c,p,r}, refer to \emph{source, condensate, probe and
response}, respectively.

In general the relevant parts of the $S_N$'s are of the forms:\bea
\begin{split}S_N^{(1)}&=\int{...+\;c_B\;\delta\tilde{\phi}_p\;\tilde{\phi}_c}\;+...\cr
S_N^{(2)}&=\int{...+\;c_r\;\delta\tilde{\phi}_p\;\delta\tilde{\phi}_r}\;+...\end{split}\eea
where cs are constants. Then the one point function for the
operator dual to $\phi$ will be:\be\langle O\rangle=\frac{\delta
S_{on-shell}}{\delta\tilde{\phi}_p}=c_B\;\tilde{\phi}_c+c_r\;\delta\tilde{\phi}_r=\langle
O\rangle_B+\langle O\rangle_r,\ee where subscripts B and r refer
to the background contribution to the v.e.v of O, and the response
of the system to the probe, respectively.

Doing the above calculation for the Einstein-Yang-Mills-Higgs
action leads to:\bea\begin{split} S(A_B+\delta
A,\phi_B+\delta\phi)=&S^{(0)}(A_B,\phi_B)+\cr
&S_{\phi}^{(1)}(A_B,\phi_B,\delta\phi)+S_{A}^{(1)}(A_B,\phi_B,\delta
A)+\cr
&S_{\phi}^{(2)}(A_B,\phi_B,\delta\phi^2)+S_{A\phi}^{(2)}(A_B,\phi_B,\delta
A,\delta\phi)+S_{A}^{(2)}(A_B,\phi_B,\delta A^2)+...\cr
=&S^{(0)}(A_B,\phi_B)+\cr&S_{N\phi}^{(1)}(A_B,\phi_B,\delta\phi)+S_{NA}^{(1)}(A_B,\phi_B,\delta
A)+\cr
&S_{N\phi}^{(2)}(A_B,\phi_B,\delta\phi^2)+S_{NA}^{(2)}(A_B,\phi_B,\delta
A^2)+...;\end{split}\eea In the last line we have used the
equations of motion for background and linear equations for
fluctuations. For Einstein-Yang-Mills-Higgs action, the Noether
actions
are:\bea\begin{split}S^{(1)}_{N\phi}&=-\int{\sqrt{-g}\;\delta\vec{\phi}.D^\mu\vec{\phi_B}\;ds_\mu}=-\int{\sqrt{-g}\;\delta\vec{\phi}.D^r\vec{\phi_B}\;d^3x}\cr
S^{(2)}_{N\phi}&=-\frac{1}{2}\int{\sqrt{-g}\;\delta\vec{\phi}.D^\mu\delta\vec{\phi}\;ds_\mu}=-\frac{1}{2}\int{\sqrt{-g}\;\delta\vec{\phi}.D^r\delta\vec{\phi}\;d^3x}\cr
S^{(1)}_{NA}&=-\int{\sqrt{-g}\;\delta\vec{A}_{\nu}.\vec{F}_B^{\mu\nu}\;ds_\mu}=-\int{\sqrt{-g}\;\delta\vec{A}_{\nu}.\vec{F}_B^{r\nu}\;d^3x}\cr
S^{(2)}_{NA}&=-\frac{1}{2}\int{\sqrt{-g}\;\delta\vec{A}_{\nu}.\delta\vec{F}^{\mu\nu}\;ds_\mu}=-\frac{1}{2}\int{\sqrt{-g}\;\delta\vec{A}_{\nu}.\delta\vec{F}^{r\nu}\;d^3x}\end{split}\eea
where $ds_\mu$ is the area element normal to the boundary.

So roughly\footnote{By "$\propto$", we mean the v.e.vs are proportional to one of the expansion coefficients which couples to the probe, as mentioned previously.} we have:\be \langle
\vec{V}_\mu\rangle_B\propto(\vec{F}_B)_{r\mu}\;,\;\langle
\vec{O}\rangle_B\propto\partial_r\vec{\phi}_B.\ee

The v.e.v of the charge density operator ($V_0=\rho$), is
proportional to the radial electric field; the v.e.v of spatial
components of the vector operator ($V_i$), are proportional to the
\emph{tangent to the boundary components} of the magnetic field, and the v.e.v of the
scalar operator is proportional to the radial derivative of the
scalar field. So for Enstein-Yang-Mills-Higgs
action, the \emph{radial magnetic field}, which is the
characteristic of magnetic monopoles, doesn't contribute to v.e.vs as a condensate; but, its boundary value plays the role of a source for the dual field theory. If we want the radial magnetic field to contribute to the v.e.vs, as a condensate, we have to add other terms to the action. For example adding a
$\theta$-term leads to:\bea\begin{split}
\langle\vec{\rho}\rangle&\propto \vec{E}_r+\theta\;\vec{B}_r\cr
\langle\vec{V}_i\rangle&\propto
\epsilon_{ij}\;(\vec{B}_j+\theta\;\vec{E}_j)\end{split}\eea
\subsection{Julia-Zee dyon and Vortex condensation}
\begin{figure}
  \centering
  \subfloat[]{\includegraphics[scale=.3]{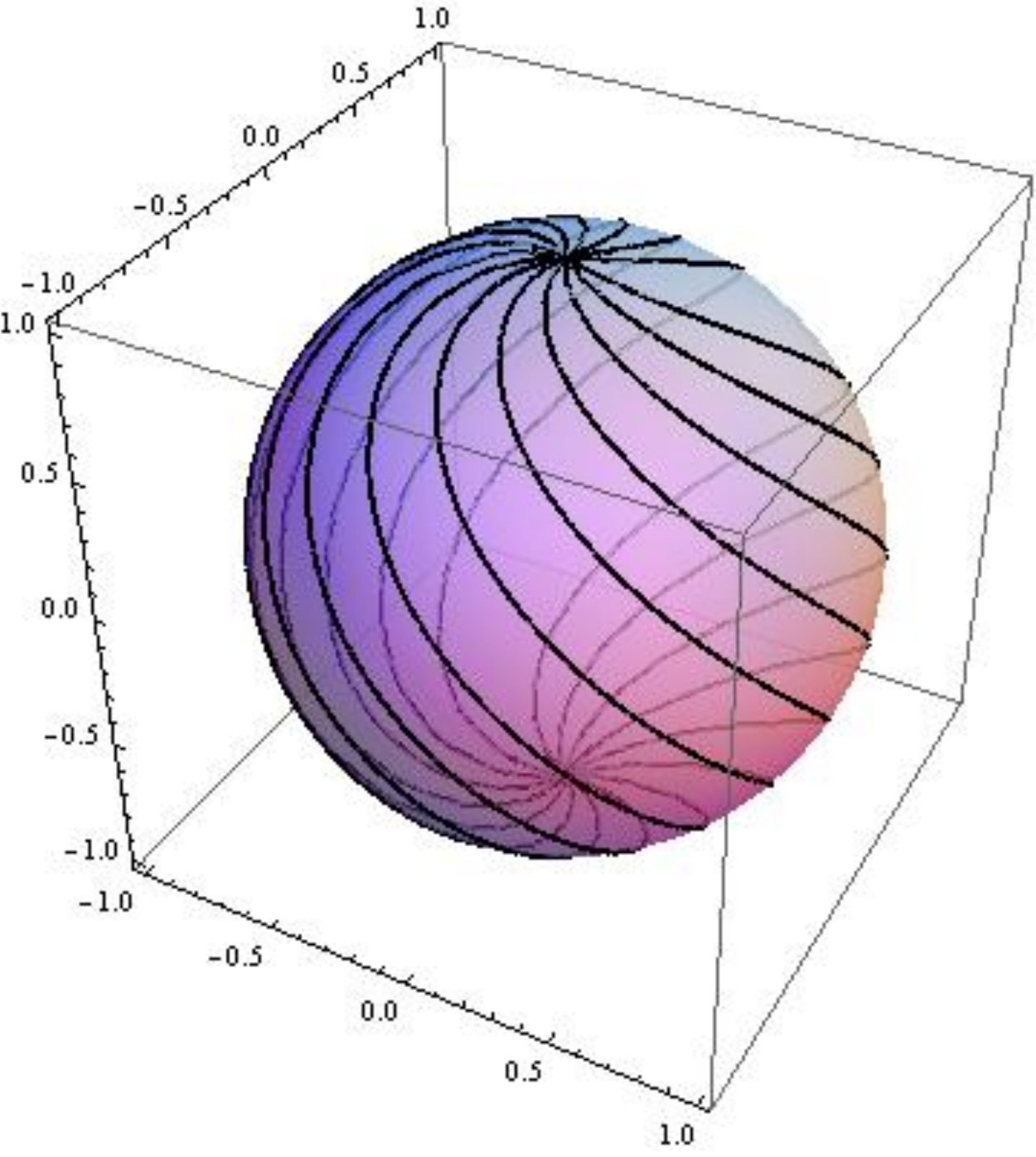}}
  \subfloat[]{\includegraphics[scale=.3]{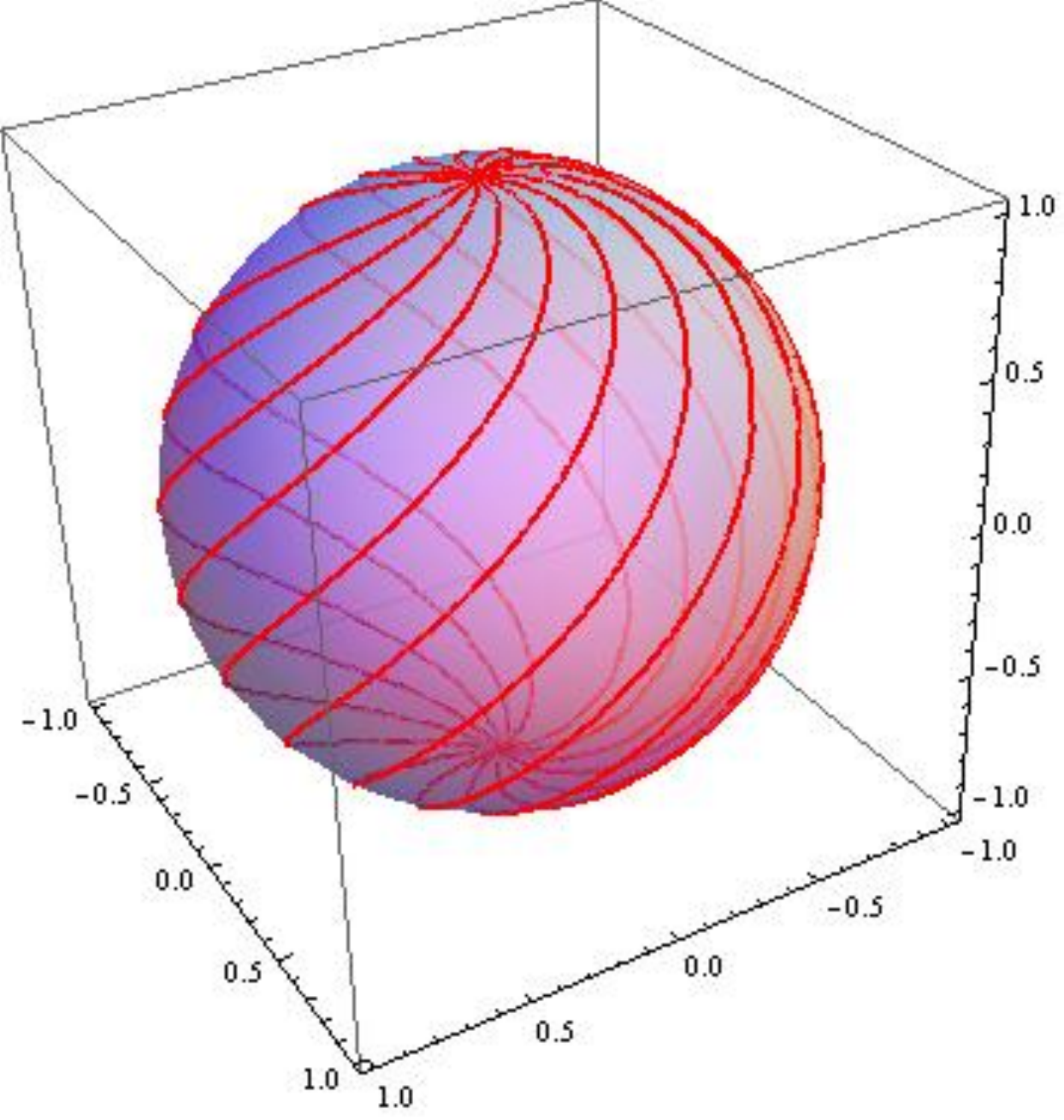}}
  \caption{plot of directed curves for $e_+$ (a) and $e_-$ (b) basis.}
\end{figure}

Applied to the Julia-Zee dyon, it is easy to see that: \bea
\langle \vec{V}_t\rangle\propto(\vec{E})_{r}\propto
(\frac{J}{r})'\;\hat{r}\cr\langle
\vec{V}_i\rangle\propto\epsilon_{ij}(\vec{B})_{j}\propto
K'\;\epsilon_{ij}\partial_j\hat{r}\cr\langle
\vec{O}\rangle\propto\partial_r\vec{\phi}\propto
(\frac{H}{r})'\;\hat{r}.\eea Thus for both solutions in (2.7),
these quantities are zero, except for the charge density. In the next
section we will introduce a numerical set of solutions for which
the vacuum expectation value of the operators are non zero, and we
will study them in the phase space; but, before we proceed, it is
interesting to look at the profile of the vector condensate.

By looking at the ansatz (2.4), we see that only two components of
the vector field are non zero, $A_\theta^\varphi$ and
$A_\varphi^\theta$; where the subscripts refer to spatial indices
and superscripts are internal indices. Projecting this vector on
the surface, by stereographic projection, we see that the profile
of the vector is that of a \emph{vortex}. These vectors do not have
definite charge under $\sigma^r$; but, after some simple algebra we can
change the basis to have definite charge vectors. As usual we
can define $\tau^\pm=\tau^\theta\pm i\tau^\varphi$ and write\be
A=\frac{2J}{er}\;\tau^r
\;dt+\;\frac{1-K}{e}\big[\tau^+\;\hat{e}_++\tau^-\;\hat{e}_-\big]\ee
where the $\hat{e}_\pm$ are:\be
\hat{e}_\pm=sin(\theta)\;d\varphi\;\pm\;i\;d\theta. \ee As it is
clear, by multiplying these basis with a length factor R, the real
part of the basis is the azimuthal and the imaginary part of the
basis is radial differential length elements in the complex plane.
In figure (1) we have shown the directed curves for $e^\pm$ basis.
At the poles the basis is completely polar, and at the equator
there are equal contributions to the $\hat{e}_\pm$, from $\theta$
and $\varphi$ coordinates. The profile of the vector in this basis
is also similar to a vortex. In hedgehog gauge the magnetic charge
is related to $\pi_2(S^2)$, but, in abelian gauge it is the magnetic flow
which is carried by the Dirac string\cite{Arafune:1975ab}; since the Dirac string passes through the center of the vortex, so the
magnetic charge is the magnetic flow which is trapped by the
vortex. In figure (2), we have shown the profile of the
vector condensate, in hedgehog gauge.
\begin{figure}
  \centering
  \includegraphics[scale=.5]{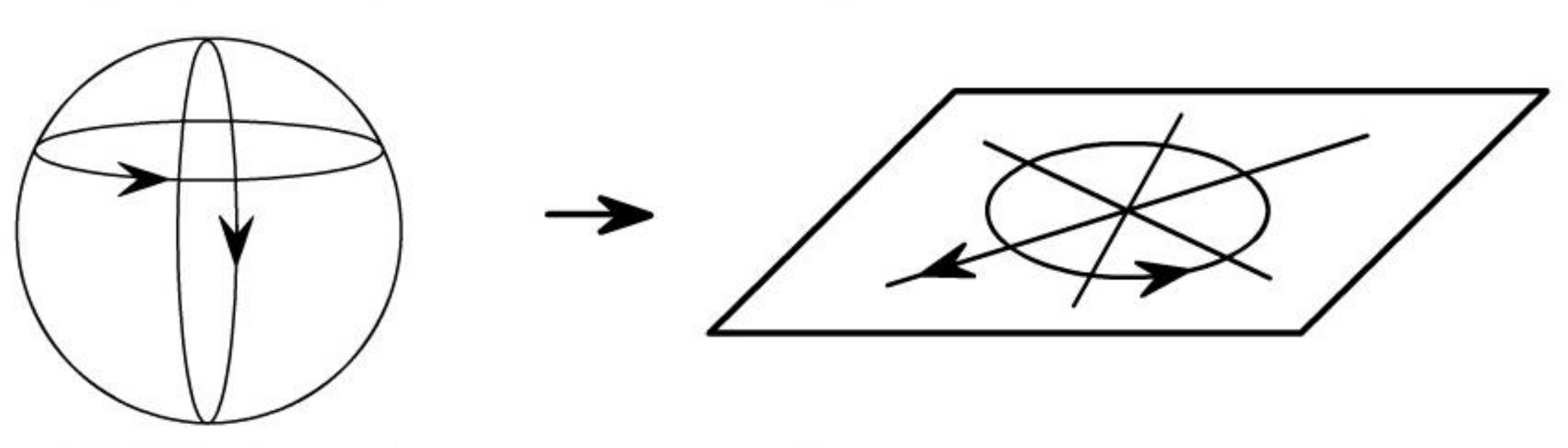}
  \caption{profile of the vector condensate.}
\end{figure}

\section{Numerical calculations}
In this section we construct a numerical self-garvitating
Julia-Zee dyon, and study the behavior of the v.e.vs as functions
of the sources of the model. The equation of motion for the matter
functions are:\bea
\begin{split}&J^{\prime\prime}-\frac{r}{2}(X+Y)^{\prime}\big(\frac{J}{r}\big)^\prime-e^Y\frac{2JK^2}{r^2}=0,
\cr &K^{\prime\prime}+\frac{1}{2}(X-Y)^\prime
K^\prime-e^Y\frac{K}{r^2}\big(K^2+H^2-e^{-X}J^2-1\big)=0, \cr
&H^{\prime\prime}+\frac{r}{2}(X-Y)^\prime\big(\frac{H}{r}\big)^\prime-e^Y\frac{H}{r^2}\big(2K^2+\frac{\lambda}{e^2}(H^2-c^2r^2)\big)=0;\end{split}
\eea and the equations for metric functions are:\bea
\begin{split}&\frac{e^{-Y}}{r^2}(rY^\prime-1)+\frac{1}{r^2}+\frac{3}{L^2}=\frac{8\pi
G}{e^2}T^t_t \cr
-&\frac{e^{-Y}}{r^2}(rX^\prime+1)+\frac{1}{r^2}+\frac{3}{L^2}=\frac{8\pi
G}{e^2}T^r_r \cr
-&\frac{e^{-Y}}{2}\big[X^{\prime\prime}+\frac{1}{2}(X^\prime)^2-\frac{1}{2}X^\prime
Y^\prime+\frac{1}{r}(X-Y)^\prime\big]+\frac{3}{L^2}=\frac{8\pi
G}{e^2}T^\theta_\theta=\frac{8\pi
G}{e^2}T^\varphi_\varphi,\end{split}\eea where \bea
\begin{split}
&T^t_t\;=\;\bigg[\frac{e^{-Y}}{r^2}(K^\prime)^2+\frac{(K^2-1)^2}{2r^4}+\frac{e^{-(X+Y)}}{2}[(\frac{J}{r})^\prime]^2+\cr
&e^{-X}\frac{J^2K^2}{r^4}+\frac{e^{-Y}}{2}[(\frac{H}{r})^\prime]^2+\frac{H^2K^2}{r^4}+e^2
V(\phi)\bigg]\cr\cr
&T^r_r\;=\;\bigg[-\frac{e^{-Y}}{r^2}(K^\prime)^2+\frac{(K^2-1)^2}{2r^4}+\frac{e^{-(X+Y)}}{2}[(\frac{J}{r})^\prime]^2-\cr
&e^{-X}\frac{J^2K^2}{r^4}-\frac{e^{-Y}}{2}[(\frac{H}{r})^\prime]^2+\frac{H^2K^2}{r^4}+e^2
V(\phi)\bigg]\cr\cr
&T^\theta_\theta\;=\;T^\varphi_\varphi\;=\;\bigg[-\frac{(K^2-1)^2}{2r^4}-\frac{e^{-(X+Y)}}{2}[(\frac{J}{r})^\prime]^2+\frac{e^{-Y}}{2}[(\frac{H}{r})^\prime]^2+e^2
V(\phi)\bigg].
\end{split}\eea To find a solution we assume that there is a horizon at
$r_H$, and consider a solution with the near horizon expansion of
the form:\bea\begin{split} J(r)&=J_1(r-r_H)+J_2(r-r_H)^2+...\cr
H(r)&=H_0+H_1(r-r_H)+H_2(r-r_H)^2+...\cr
K(r)&=K_0+K_1(r-r_H)+K_2(r-r_H)^2+...\cr
e^{X(r)}&=X_1(r-r_H)+X_2(r-r_H)^2+...\cr
e^{-Y(r)}&=Y_1(r-r_H)+Y_2(r-r_H)^2+...\end{split}\eea using the
equations of motion it is easy to see that only $J_1,H_0,K_0,X_1$
are independent coefficients, and other coefficients can be
determined from them.

Our system of equations has a scaling symmetry:\bea
\begin{split} &(r,r_H,t,L)\rightarrow\alpha\; (r,r_H,t,L)\cr
&(ds^2,G_N)\rightarrow\alpha^2\;(ds^2,G_N)\cr
&c^2\rightarrow\frac{1}{\alpha^2}c^2\cr
&(\theta,\varphi,J,K,H,e^X,e^Y,\lambda,e)\rightarrow(\theta,\varphi,J,K,H,e^X,e^Y,\lambda,e)
\end{split}\eea
Thus, choosing $\alpha=1/r_H$, we can set $r_H=1$; we then
define\footnote{In fact by this rescaling all the quantities will
be in the units of $r_H$.} \bea\begin{split}x
&=\frac{r}{r_H}\cr\tilde{L}&=\frac{L}{r_H}\cr \tau
&=\frac{t}{r_H}\end{split}\eea ($\tilde{L}$ in this definition
should not be confused with $\tilde{L}$ in (2.7)). So that by
starting from these expansions and running to large radii, from
our numerical calculations, we see that the asymptotic form of the
functions are:\bea\begin{split} J(x)&=\mu x-\rho+...\cr
H(x)&=\tilde{H}_0+\frac{\tilde{H}_1}{x}+...\cr
K(x)&=\tilde{K}_0+\frac{\tilde{K}_1}{x}+...\cr
e^{X(x)}&=c^2\;\frac{x^2}{\tilde{L}^2}+...\cr
e^{-Y(x)}&=\frac{x^2}{\tilde{L}^2}+...\cr \rightarrow
d\tilde{s}^2&=\frac{1}{r_H^2}\; ds^2 \sim
-\frac{x^2}{\tilde{L}^2}\;c^2\;d\tau^2+\frac{\tilde{L}^2}{x^2}\;dx^2+x^2\;d\Omega_2^2.\end{split}\eea
Note that generally in our numerical calculations, $c\neq1$. In
order for the asymptotic speed of light to be 1, we have to
rescale the time. In fact there is another scale symmetry in the
equations of motion:\be X(r)\rightarrow X(r)+2\beta \;;\;
t\rightarrow e^{-\beta}t \;;\; A_t\rightarrow e^\beta A_t. \ee
Using this scaling we can write \be \tau\rightarrow
\tilde{t}=c\;\tau. \ee So the correct physical time is \be
\tilde{\tau}=r_H\;\tilde{t}=c\;r_H\;\tau. \ee Since what we
calculate numerically are $g_{xx},g_{\tau\tau}$ and $A_\tau$, for
calculating temperature, chemical potential and charge density
(and all quantities which are related to time coordinate in
someway) \emph{correctly}, this rescaling must be considered in
the calculations. So for the temperature we have:\be
T=\frac{1}{2\pi}\big(\;\sqrt{g^{rr}}\partial_r\sqrt{-g_{\tilde{\tau}\tilde{\tau}}}\;\big)_{r=r_H}=\frac{1}{2\pi\;c
\;r_H}\big(\;\sqrt{g^{xx}}\partial_x\sqrt{-g_{\tau\tau}}\;\big)_{x=1}.
\ee In the series considered below, we use in our numerical
calculations, \bea\begin{split} g^{xx}=Y_1(x-1)+...\cr
g_{\tau\tau}=X_1(x-1)+...
\end{split}\eea the correct temperature is:\be T=\frac{1}{4\pi\;c\; r_H }\sqrt{X_1Y_1}.\ee
Also concerning the chemical potential we have, \be
A_\tau\;d\tau=\big(\;A_\tau
\frac{1}{c\;r_H}\;\big)d\tilde{\tau}=A_{\tilde{\tau}}\;d\tilde{\tau};
\ee so that $A_{\tilde{\tau}}$ is the physical field.

In order to calculate the vacuum expectation values, we should
study the fluctuations on the background. As can be seen from
(3.7) to get the background v.e.vs it is enough to consider
the fluctuations of the same form as those of the background. So by varying
the matter functions\footnote{For simplicity we consider only
 s-wave fluctuations},\bea
\begin{split} &\frac{H}{er} \rightarrow \frac{H}{er}+\epsilon \;
\frac{h(r)}{e},\cr &\frac{J}{er}\rightarrow \frac{J}{er}+\epsilon
\; \frac{P(r)}{e},\cr &\frac{K}{e}\rightarrow \frac{K}{e}+\epsilon
\; \frac{Q(r)}{e},
\end{split} \eea and solving the linearized equations for the
fluctuations on the background of the calculated numerical
solution, with the same initial conditions as the background
functions, we see that their asymptotic forms are the same as those of the
background:\bea\begin{split}h(r)&=\frac{\tilde{h}_0}{r}+\frac{\tilde{h}_1}{r^2}+...\cr
P(r)&=\tilde{P}_0+\frac{\tilde{P}_1}{r}+...\cr
Q(r)&=\tilde{Q}_0+\frac{\tilde{Q}_1}{r}+...;\end{split}\eea so
$\mu,\tilde{H}_0,1-\tilde{K}_0$, are sources and
$\rho,\tilde{H}_1,\tilde{K}_1$, are condensates; also
$\tilde{h}_0,\tilde{P}_0,\tilde{Q}_0$, are probes and
$\tilde{h}_1,\tilde{P}_1,\tilde{Q}_1$, are responses.

\begin{figure}[H]
  \centering
  \subfloat[]{\includegraphics[scale=.7]{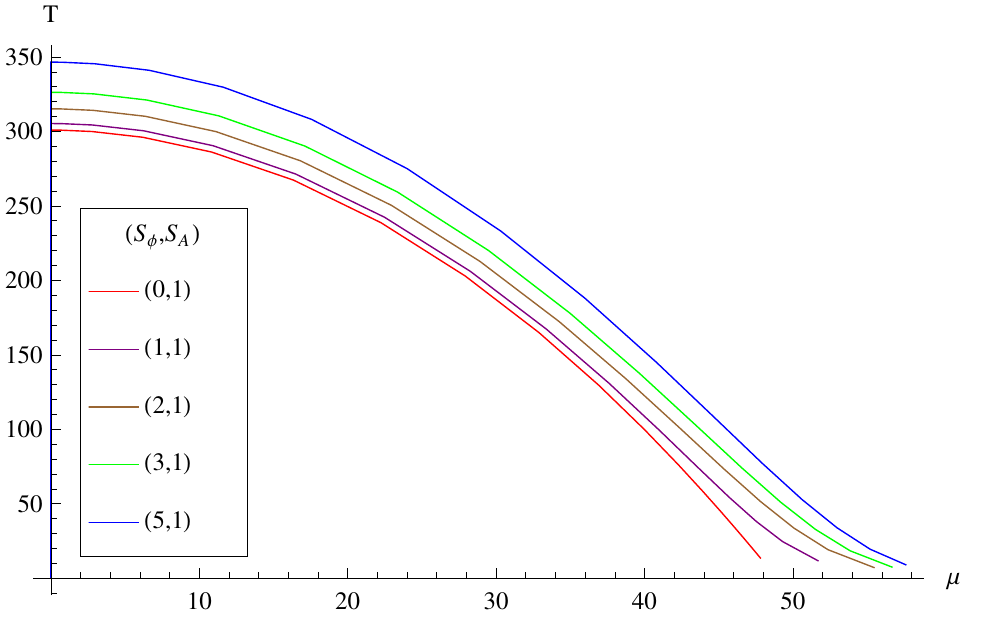}}
  \subfloat[]{\includegraphics[scale=.7]{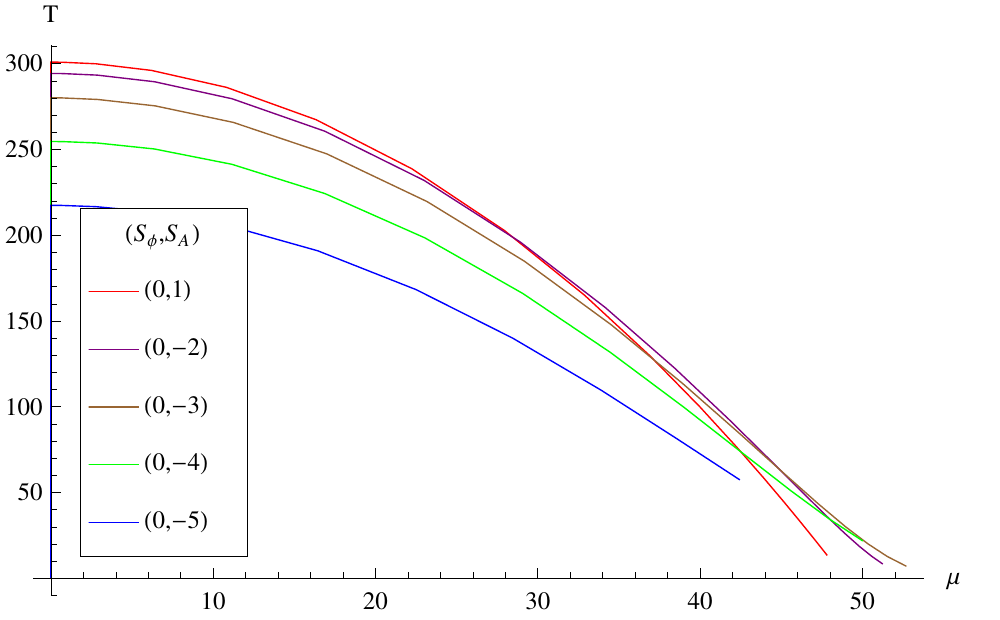}}
  \caption{plot of the phase subspace when $S_\phi$ and $S_A$ are constant. The red line is the AdS-RN solution.}
\end{figure}
When the radius of the horizon is fixed, although we have four
sources ($T,\mu,S_\phi,S_A$) in the model, the phase space is
a three dimensional hypersurface. It is similar to equation
(2.11), where temperature and chemical potential are constrained
and the phase space is one dimensional. In this case we expect the
constraint to be:\be
\frac{\epsilon+P}{s}=T+\mu\;\frac{\rho}{s}+S_\phi\;\frac{\langle
o\rangle}{s}+S_A\;\frac{\langle v\rangle}{s}. \ee In figure (3) we
have plotted some curves related to some numerical solutions in
the ($T,\mu$) plane, for fixed values of $S_\phi$ and $S_A$.

From the numerical calculations the following emerges:
\begin{enumerate}
    \item The
solutions with the desired boundary conditions exist only for a
limited range of $S_\phi$ and $S_A$ and also temperature (chemical
potential).
    \item As is shown in figures (4,5), the
vacuum expectation value of the vector operator generally has
different signs in different regions of the phase space, and there
is a two dimensional surface on which the vacuum expectation value
is zero. The sign changes smoothly. This phenomenon can be
interpreted as a vortex reversing\footnote{Note that, as is clear
from the equations of motion, the absolute sign of the matter
functions (J,H,K) is not important, although sign changing is
meaningful. In this paper we have considered positive solutions,
(J,H,K)$>$ 0.}.
    \item At $\mu=S_\phi=0$, the sign of $\langle V \rangle$
    changes at $(S_A)_{cr}=0$, or alternatively,
    $(\tilde{K}_0)_{cr}=1$. Since the radial
    nonabelian magnetic field is proportional to $
    \vec{B}_r\propto (1-K(r)^2)\hat{r}$, the boundary magnetic
    field is $(\vec{B}_r)_{boundary}\propto (1-\tilde{K}_0^2)\hat{r}$. So at $\mu=S_\phi=0$,
    the vortex reverses when the boundary magnetic field is reversed.
    \item As is clear from figures (4,5), at nonzero chemical potential,
    even when the boundary magnetic field is zero, the vacuum expectation value of the vector operator can be nonzero; So the vortex can be a condensate.
    In order to cancel the vortex, we should impose a boundary magnetic field.
    Increasing $\mu$, increases $(\tilde{K}_0)_{cr}$ and alternatively $(\vec{B}_r)_{boundary}$ which is needed to cancel the vortex;and increasing $S_\phi$, decreases $(\tilde{K}_0)_{cr}$.
    \item When $S_\phi$ is small enough, similar to the vector operator, the vacuum expectation value
of the scalar operator has different signs in different regions of
the phase space, and there is a two dimensional surface on which
the vacuum expectation value is zero, but this time it occurs
at low temperatures. This phenomenon can be related to an
instability, because in\cite{Allahbakhshi:2010ii}, it is shown that such an event
occurs for linear fluctuations which means that at some special
value of parameters the equations admit a zero mode solution, as a
marginal stable mode. For vector fluctuations this is not the
case.
\end{enumerate}

\begin{figure}
  \centering
  \subfloat[]{\includegraphics[scale=.7]{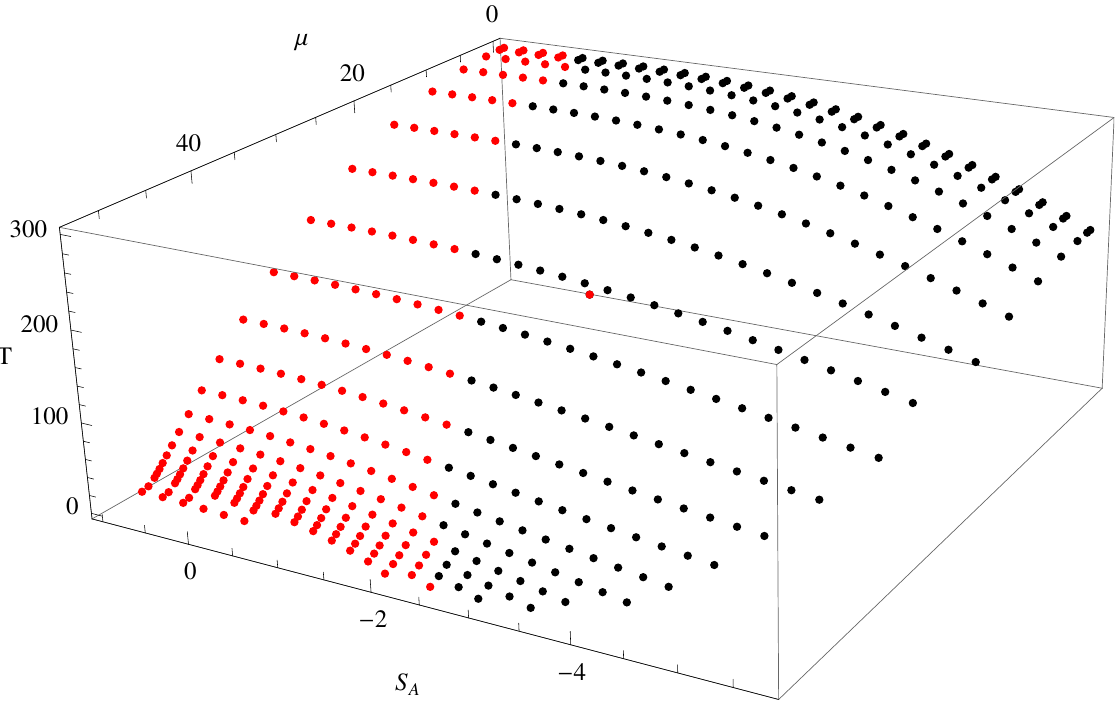}}
  \subfloat[]{\includegraphics[scale=.7]{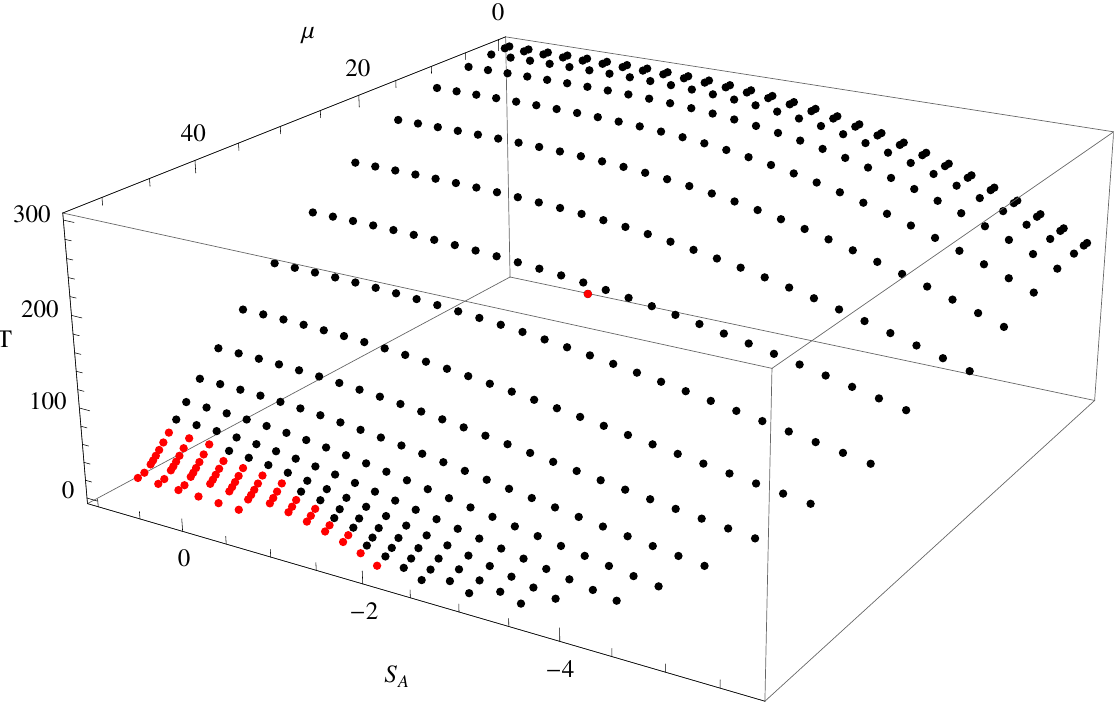}}
  \caption{plot of the phase subspace when $S_\phi=0.1$. The red region is where the v.e.vs are negative, and in black region they are positive. Plot (a) is for $\langle V\rangle$, and plot (b) is for $\langle O\rangle$}
\end{figure}

\begin{figure}[]
  \centering
  \subfloat[]{\includegraphics[scale=.58]{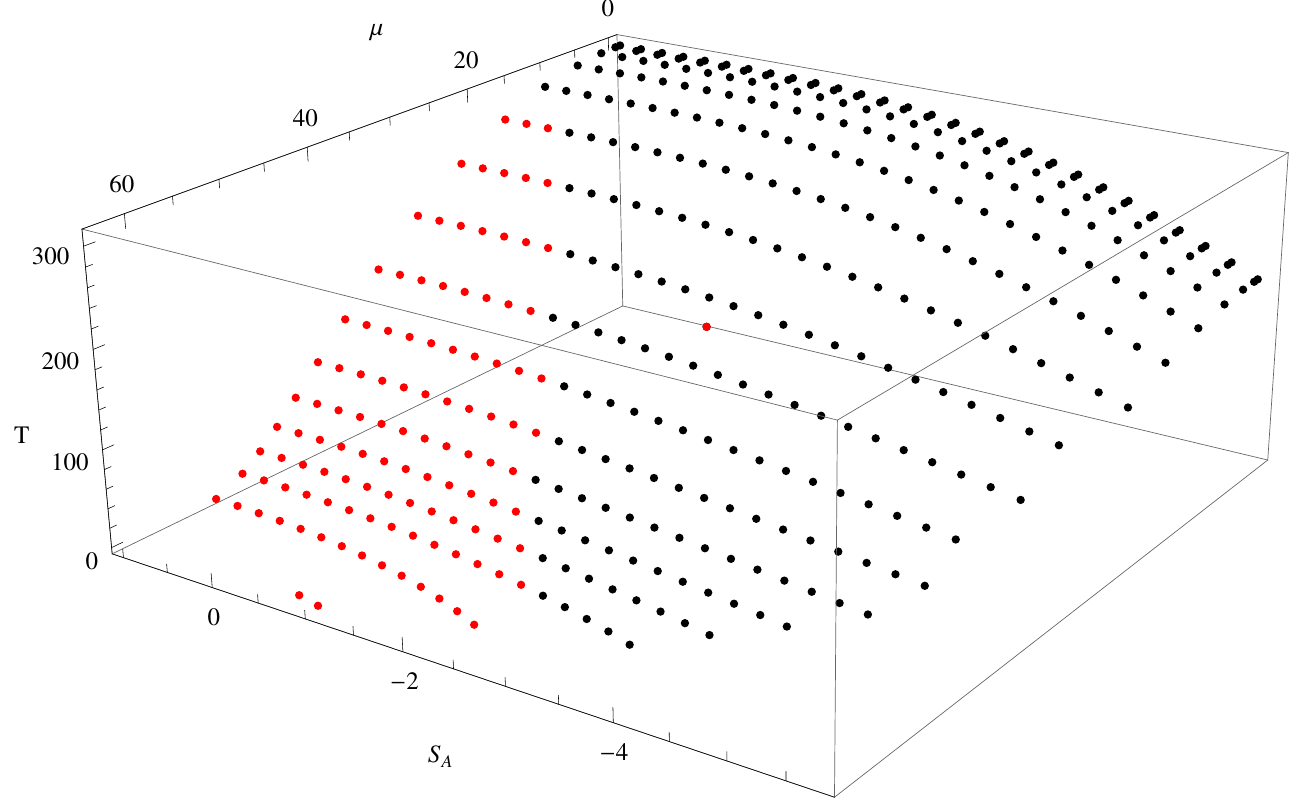}}
  \subfloat[]{\includegraphics[scale=.58]{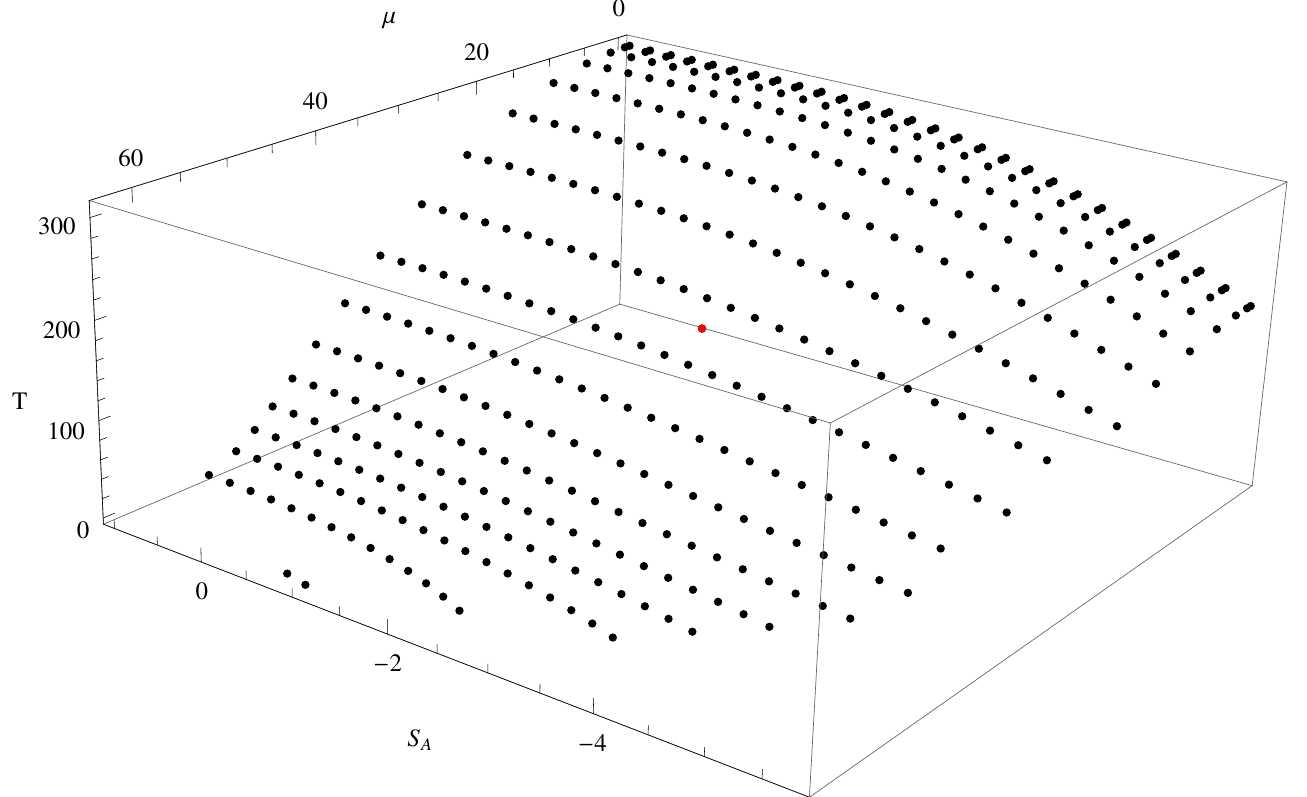}}
  \caption{plot of the phase subspace when $S_\phi=1.5$. Plot (a) is for $\langle V\rangle$, and plot (b) is for $\langle O\rangle$. The negative region for $\langle O\rangle$ is disappeared.}
\end{figure}

\section{Conclusion}

In this article we studied the vacuum expectation values of dual
operators from Einstein-Yang-Mills-Higgs action in a holographic
setup and found that the radial magnetic field which is the main
property of the magnetic monopoles does not play explicit role in
the vacuum expectation values. In particular we consider the
Julia-Zee dyon, and saw that the v.e.v of dual scalar and
vector operators can be nonzero, and the profile of the vector
condensate is similar to a vortex. We also constructed a numerical
self-gravitating dyon and studied the changing of the v.e.vs as a
function of the sources. We observed a sign changing phenomenon in the
v.e.vs by changing temperature, chemical potential, and scalar and
vector sources. It will be interesting to check whether these
sign changings relate to an instability. Also it seems reasonable
to expect that the multi monopole solutions which are studied
in\cite{Bolognesi:2010nb}, correspond to multi vortex
configurations in the dual field theory side. Any firm statement
on this requires knowledge of the profile of the vector field in the
bulk.
\\\\\\
\textbf{Acknowledgements} \\I would like to thank Farhad Ardalan,
for extensive discussions and reading the manuscript, M. Alishahiha and A.E. Mosaffa and also R.
Fareghbal, A. Davody, A. Naseh and R. Mozaffari for useful
discussions.


\begin{thebibliography}{}

\bibitem{Maldacena:1997re}
  J.~M.~Maldacena,
  ``The large N limit of superconformal field theories and supergravity,''
  Adv.\ Theor.\ Math.\ Phys.\  {\bf 2}, 231 (1998)
  [Int.\ J.\ Theor.\ Phys.\  {\bf 38}, 1113 (1999)]
  [arXiv:hep-th/9711200].

\bibitem{Witten:1998qj}
  E.~Witten,
  ``Anti-de Sitter space and holography,''
  Adv.\ Theor.\ Math.\ Phys.\  {\bf 2}, 253 (1998)
  [arXiv:hep-th/9802150].

\bibitem{Gubser:1998bc}
  S.~S.~Gubser, I.~R.~Klebanov and A.~M.~Polyakov,
  ``Gauge theory correlators from non-critical string theory,''
  Phys.\ Lett.\  B {\bf 428}, 105 (1998)
  [arXiv:hep-th/9802109].

\bibitem{Lugo:2010qq}
  A.~R.~Lugo, E.~F.~Moreno and F.~A.~Schaposnik,
  ``Holographic phase transition from dyons in an AdS black hole background,''
  JHEP {\bf 1003}, 013 (2010)
  [arXiv:1001.3378 [hep-th]].

\bibitem{Lugo:2010ch}
  A.~R.~Lugo, E.~F.~Moreno, F.~A.~Schaposnik,
  ``Holography and $AdS_4$ self-gravitating dyons,''
  JHEP {\bf 1011}, 081 (2010).
  [arXiv:1007.1482 [hep-th]].

\bibitem{Allahbakhshi:2010ii}
  D.~Allahbakhshi, F.~Ardalan,
  ``Holographic Phase Transition to Topological dyons,''
  JHEP {\bf 1010}, 114 (2010).
  [arXiv:1007.4451 [hep-th]].

\bibitem{vanderBij:2002sq}
  J.~J.~van der Bij, E.~Radu,
  ``Magnetic charge, angular momentum and negative cosmological constant,''
  Int.\ J.\ Mod.\ Phys.\  {\bf A18}, 2379-2393 (2003).
  [hep-th/0210185].

\bibitem{Radu:2004ys}
  E.~Radu, D.~H.~Tchrakian,
  ``New axially symmetric Yang-Mills-Higgs solutions with negative cosmological constant,''
  Phys.\ Rev.\  {\bf D71}, 064002 (2005).
  [hep-th/0411084].

\bibitem{Bolognesi:2010nb}
  S.~Bolognesi, D.~Tong,
  ``Monopoles and Holography,''
  JHEP {\bf 1101}, 153 (2011).
  [arXiv:1010.4178 [hep-th]].

\bibitem{Sutcliffe:2011sr}
  P.~Sutcliffe,
  ``Monopoles in AdS,''. [arXiv:1104.1888 [hep-th]].


\bibitem{Kasuya:1981tq}
  M.~Kasuya and M.~Kamata,
  ``An Exact Dyon Solution With The Reissner-Nordstrom Metric,''
  Nuovo Cim.\  B {\bf 66}, 75 (1981).


\bibitem{Arafune:1975ab}
  J.~Arafune, P.~G.~O.~Freund and C.~J.~Goebel,
  ``Topology Of Higgs Fields,''
{\it  IN *Kyoto 1975, Proceedings. Lecture Notes in Physics*,
Berlin 1975, 240-241}


\end{thebibliography}
\end{document}